\documentclass[a4paper]{jpconf}
\usepackage{graphicx}
\begin{document}
\title{Field-Induced Quantum Spin Nematic Liquid Phase in the S=1 Antiferromagnetic Heisenberg Chain with Additional Interactions }

\author{T\^oru Sakai$^{1,2}$, Hiroki Nakano$^1$, Rito Furuchi$^1$ and Kiyomi Okamoto$^1$}

\address{
Graduate School of Science, University of Hyogo, Hyogo 678-1297, 
Japan 
}

\ead{sakai@spring8.or.jp}

\begin{abstract}

The magnetization process of the $S=1$ antiferromagnetic chain with the single-ion anisotropy $D$ and the biquadratic interaction is investigated using the 
numerical diagonalization. Both interactions stabilize the 2-magnon Tomonaga-Luttinger liquid (TLL) phase in the magnetization process. 
Based on several excitation gaps calculated by the numerical diagonalization, some phase diagrams of the magnetization process are 
presented. These phase diagrams reveal that the spin nematic dominant TLL phase appears at higher magnetizations for sufficiently 
large negative $D$.

\end{abstract}

\section{Introduction}

The quantum spin nematic state is one of interesting topics in the 
field of the strongly correlated electron systems. 
It was theoretically predicted to be realized in 
several frustrated systems; the square-lattice model with the 
ferromagnetic nearest-neighbor and antiferromagnetic next-nearest-neighbor 
interactions\cite{shannon}, the triangular-lattice antiferromagnetic with multi spin 
exchange interactions\cite{momoi}, and the ferromagnetic and antiferromagnetic 
zigzag chain\cite{chubukov,hikihara}. 
The high-magnetic field measurement\cite{nawa} on the zigzag chain compound 
LiCuVO$_4$ detected an anomalous behavior of the magnetization curve 
just below the saturation magnetization, which was supposed to be the 
spin nematic phase.  The quantum spin liquid like behavior of the 
$S=1$ triangular-lattice compound NiGa$_2$S$_4$ was theoretically 
explained as the spin nematic state\cite{nakatsuji}.  
A density matrix renormalization group (DMRG) analysis\cite{mila} 
indicated that the field-induced spin nematic liquid phase appears 
in the $S=1$ antiferromagnetic chain with the biquadratic 
interaction. 
It was supported by the numerical diagonalization study\cite{sakain}.

On the other hand, the 2-magnon bound state similar to the spin 
nematic state was predicted to appear in the magnetization process 
of the $S=1$ antiferromagnetic chain with the easy axis anisotropy 
based on the numerical diagonalization analysis\cite{sakai1}. 
It would possibly correspond to the spin nematic state. 
In the present paper, we investigate the $S=1$ antiferromagnetic 
chain with the easy axis anisotropy and the biquadratic interaction 
using the numerical diagonalization, 
in order to consider the relation between this 2-magnon 
bound state and the spin nematic liquid phase. 

\section{Model}

We consider the magnetization process of 
the $S=1$ antiferromagnetic chain with the easy axis single ion anisotropy 
$D$ and the biquadratic interaction $J_{\rm BQ}$. 
The system under the magnetic field is described by the Hamiltonian
\begin{eqnarray}
{\cal H}&=&{\cal H}_0+{\cal H}_Z \cr 
{\cal H}_0&=&J\sum_{j=1}^L \vec{S}_j\cdot \vec{S}_{j+1} 
+ D \sum_{j=1}^L (S_j^z)^2
+J_{\rm BQ}\sum_{j=1}^L (\vec{S}_j\cdot \vec{S}_{j+1} )^2 \cr
{\cal H}_Z&=& -H\sum _{j=1}^L S_j^z. 
\label{ham}
\end{eqnarray}
In this paper we consider the case of $J_{\rm BQ}<0$, $D<0$ and fix $J=1$. 
The first term of ${\cal H}_0$ is the bilinear exchange interaction which 
stabilizes the antiparallel configuration of the nearest neighbor spin pair. 
On the other hand, the third term, the biquadratic interaction does 
not only stabilize the antiparallel configuration, but also the parallel 
one. Thus the spin nematic liquid phase would possibly occur for sufficiently 
large $|J_{\rm BQ}|$. 
A numerical diagonalization and DMRG  
study indicated that the biquadratic interaction stabilizes the 
spin nematic correlation\cite{lauchli}. 
If the spin nematic liquid phase is realized, 
the four spin correlation function would exhibit the power-law decay, 
while the two spin one decays exponentially. 
Then in this phase the single magnon excitation should be gapped, 
but the 2-magnon excitation should be gapless. 
Therefore the 2-magnon bound state would be realized there. 
If the spin nematic liquid phase appears in the magnetization 
process described by the Hamiltonian (\ref{ham}), 
each step of the magnetization curve for the finite-size systems 
should be $\delta S^z_{\rm total} =2$, 
while $\delta S^z_{\rm total} =1$ in the conventional Tomonaga-Luttinger 
liquid (TLL) phase. 
The negative $D$ also yields a similar 2-magnon bound state, 
because the $S_j^z=\pm 1$ states are stabilized and the $S_j^z=0$ 
state is skipped\cite{sakai1}. 
In order to clarify the feature of these 2-magnon bound states, 
we investigate the model (\ref{ham}) using the numerical diagonalization 
of finite clusters under the periodic boundary condition. 
The calculated lowest energy for each magnetization $M=(\sum_{j=1}^LS_j^z)$ 
and each momentum $k$ is denoted as $E_k(M)$ in the following sections. 

\section{Spin Nematic Liquid}

The negative biquadratic interaction stabilizes the TLL phase where the  
2-magnon bound state is realized, as well as the negative $D$. 
In this 2-magnon TLL phase the spin correlation functions are expected to 
have the following asymptotic forms:
\begin{eqnarray}
\langle (S_0^+)^2(S_r^-)^2 \rangle \sim r^{-{\eta}_2}, \quad
\langle S_0^z S_r^z \rangle \sim \cos(2k_F r)r^{-{\eta}_z} \quad 
(r \rightarrow \infty),
\label{nematic}
\end{eqnarray}
where $2k_F$ is given by $2k_F=(1-M/M_s)\pi$ and $M_s$ is the 
saturation magnetization ($M_s=L$). 
$\langle (S_0^+)^2(S_r^-)^2 \rangle$ is the nematic (quadrapole) 
correlation perpendicular to 
the magnetic field 
and $\langle S_0^z S_r^z \rangle$ is the SDW correlation parallel to it. 
In a DMRG work\cite{mila} the region with ${\eta}_2 < {\eta}_z$ is called 
the nematic phase and the region with ${\eta}_2 > {\eta}_z$ is 
called the SDW phase, although the boundary between them is just 
a crossover line. 
Thus in order to find the spin nematic liquid phase, we should 
look for the parameter region where the 2-magnon TLL is realized 
and ${\eta}_2 < {\eta}_z$ is satisfied. 

\section{Exciation Gaps}

\begin{figure}
\includegraphics[width=0.95\linewidth,angle=0]{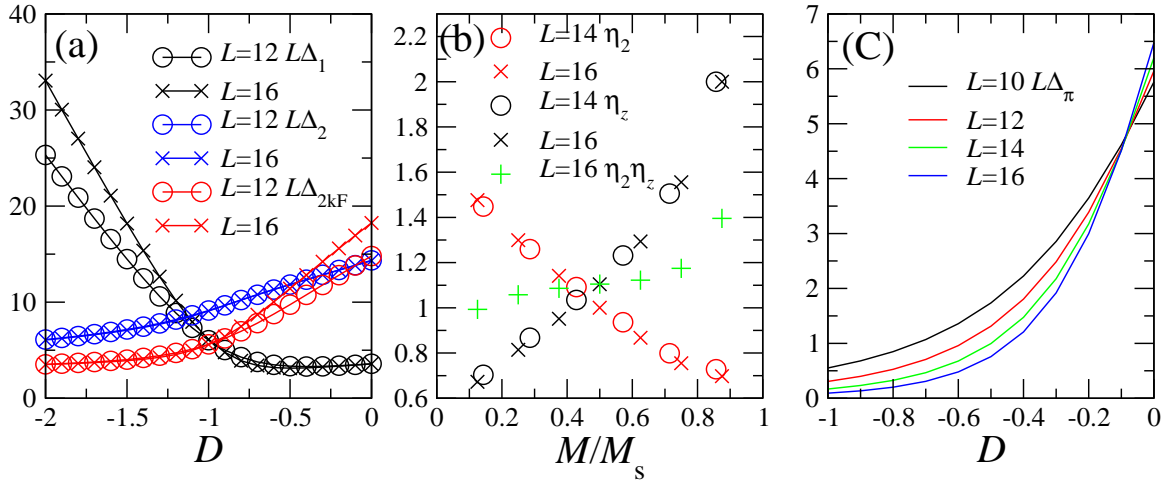}%
\caption{\label{gaps}
(a)Scaled gaps $L\Delta_1$, $L\Delta_2$ and $L\Delta_{2k_F}$ at $M=M_s/2$ 
for $J_{\rm BQ}$=-0.2 are 
plotted versus $D$ for $L=12$ and 16, 
(b)Exponents $\eta_2$ and $\eta_z$ estimated by (\ref{exponent}) 
plotted versus $M/M_s$ for $L=$14 and 16 ($J_{\rm BQ}$=-0.2, $D$=-1.5). 
$\eta_2\eta_z=1$ is well satisfied around the cross 
of $\eta_2$ and $\eta_z$. 
(c)Scaled gaps $L\Delta_{\pi}$ for $J_{\rm BQ}=-0.2$. 
It indicates the phase boundary about $D\sim -0.1$.  
 }
\bigskip
\end{figure}

In order to determine the boundary between the 1-magnon (conventional) 
and 2-magnon TLL phases, we consider several excitation gaps. 
The 2-magnon excitation gap $\Delta _2$ is gapless in both phases. 
The 1-magnon excitation gap $\Delta_1$ (the $2k_F$ excitation gap of 
the 2-magnons $\Delta_{2k_F}$) is gapless (gapped ) in the 1-magnon TLL phase, 
while gapped (gapless) in the 2-magnon one. 
$D$ dependence of the scaled excitation gaps $L\Delta_1$, $L\Delta_2$ and $L\Delta_{2k_F}$ 
at $M=M_s/2$ for $J_{\rm BQ}=-0.2$ are shown in Fig. \ref{gaps}(a). 
Fig. \ref{gaps}(a) indicates the above behaviors of these gaps. 
The cross point of $\Delta_1$ and $\Delta_{2k_F}$ is one of 
good estimations of the phase boundary between the 1-magon 
and 2-magon TLL phases. 
Our analysis of the $L$ dependence indicates that 
this phase boundary converges faster than $1/L^2$ 
with respect to $L$. 
Thus we adopt this method to determine the boundary 
between the 1-magnon and 2-magnon TLL phases. 

\section{Spin Correlation Exponents}
According to the conformal field theory, the critical exponents $\eta_2$ and $\eta_z$ 
can be estimated the forms
\begin{eqnarray}
\eta_2={{E_0(M+2)+E_0(M-2)-2E_0(M)}\over{E_{k_1}(M)-E_0(M)}}, \quad 
\eta_z=2{{E_{2k_F}(M)-E_0(M)}\over{E_{k_1}(M)-E_0(M)}},
\label{exponent}
\end{eqnarray}
for each magnetization $M$, where $k_1$ is defined as $k_1=L/2\pi$. 
$\eta_2$ and $\eta_z$ estimated for $L$=14 and 16 are plotted versus $M/M_s$ 
in Fig. 1 (b). 
It suggests that the SDW correlation is dominant for small $M$, while 
the nematic one is for large $M$. 
Since the cross point of $\eta_2$ and $\eta_z$ is not so strongly dependent of $L$, 
the cross point of $L=16$ is used as the crossover point between the SDW and nematic 
TLL phases in Fig. \ref{phase}(a), (b) and (c), for $J_{\rm BQ}$=0, -0.2 and -0.5, 
respectively. 
At least around the cross point the system well holds  the relation $\eta_2 \eta_z=1$, 
which should be satisfied for TLL, as shown in Fig. 1(b). 

\section{Phase diagrams}

\begin{figure}
\includegraphics[width=0.95\linewidth,angle=0]{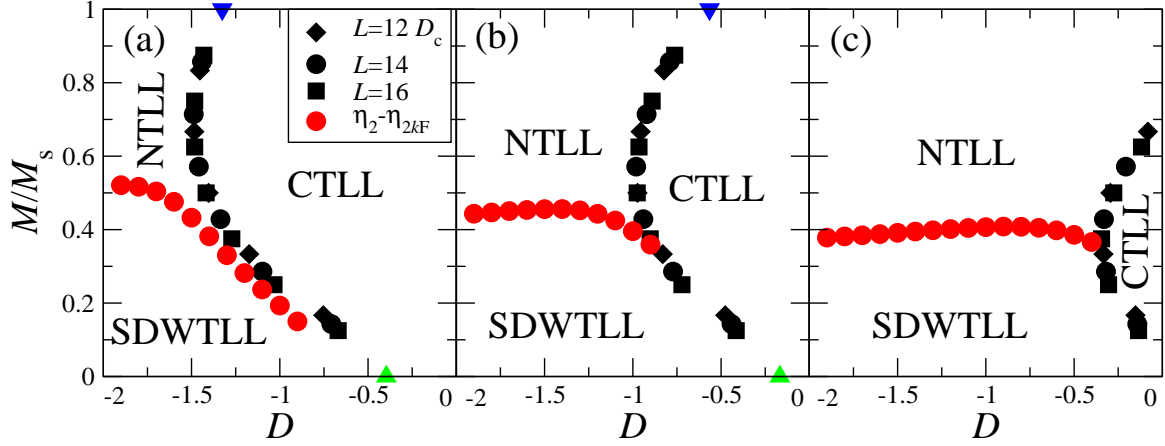}%
\caption{\label{phase}
Phase diagrams with respect to $D$ and $M/M_s$ for $J_{\rm BQ}$=0(a), -0.2(b) and -0.5(c).    
}
\bigskip
\end{figure}

The phase diagrams on the $D$-$M/M_s$ plane are 
presented for $J_{\rm BQ}$=0, -0.2, and -0.5 in Figs. \ref{phase} 
(a), (b), and (c), respectively. 
The phase boundary at $M=0$ between the Haldane and N\'eel ordered phases, 
which is denoted as the green triangle in Fig. \ref{phase},  
is determined by the phenomenological renormalization using  the 
excitation gap $\Delta _{\pi}$ with $k=\pi$ for $L$=10, 12, 14, and 16. 
The scaled gap $L\Delta_{\pi}$ is plotted versus $D$ for $J_{\rm BQ}=-0.2$ 
in Fig. \ref{gaps}(c). 
The $L$ dependent critical point obtained by $L\Delta_{\pi}(L, D_c)=(L+2)\Delta_{\pi}(L+2,D_c)$ 
can be easily extrapolated to the infinite $L$ limit. 
The critical point at $M=M_s$ is determined as the point where the 1-magnon 
and 2-magnon excitation gaps are equal, which is the blue triangle in Fig. \ref{phase}. 
Here, NTLL, SDWTLL and CTLL denote the nematic TLL, the SDW TLL, and the conventional 
TLL, respectively.
Fig.\ref{phase}(a) indicates that the NTLL can appear even for $J_{\rm BQ}=0$. 
Figs. \ref{phase}(b) and (c) suggest that the biquadratic interaction 
enhances both of NTLL and SDWTLL phases. 
The crossover boundary between NTLL and SDWTLL is always about the half the saturation 
magnetization. It is not so strongly dependent on $J_{\rm BQ}$.

\section{Summary}

The magnetization process of the $S=1$ antiferromagnetic chain with 
the easy-axis single-ion anisotropy and the biquadratic interaction 
is investigated by the numerical diagonalization for finite-size systems. 
It is found that the nematic TLL phase appears in higher magnetization 
region for sufficiently large negative $D$, even for $J_{\rm BQ}=0$. 

\section*{Acknowledgments}

This work was partly supported by JSPS KAKENHI, 
Grant Numbers  JP20K03866, 
JP18H04330 (J-Physics) and JP20H05274.
A part of the computations was performed using
facilities of the Supercomputer Center,
Institute for Solid State Physics, University of Tokyo,
and the Computer Room, Yukawa Institute for Theoretical Physics,
Kyoto University.

\end{document}